\documentclass[twocolumn]{aastex63}

\received{February 11, 2022}
\revised{March 28, 2022}
\accepted{March 29, 2022}
\shorttitle{LADUMA: Discovery of a luminous OH megamaser at $z > 0.5$}
\shortauthors{Glowacki et al.}

\begin{document}

\title{LADUMA: Discovery of a luminous OH megamaser at $z > 0.5$}

\correspondingauthor{Marcin Glowacki}
\email{marcin.glowacki@curtin.edu.au}

\author[0000-0002-5067-8894]{Marcin Glowacki}
\affiliation{Inter-University Institute for Data Intensive Astronomy (IDIA)}
\affiliation{Department of Physics and Astronomy, University of the Western Cape, Robert Sobukwe Road, Bellville 7535, South Africa}
\affiliation{International Centre for Radio Astronomy Research (ICRAR), Curtin University, Bentley, WA 6102, Australia}

\author[0000-0002-2326-7432]{Jordan D. Collier}
\affiliation{Inter-University Institute for Data Intensive Astronomy (IDIA)}
\affiliation{Department of Astronomy, University of Cape Town, Private Bag X3, Rondebosch 7700, South Africa}
\affiliation{School of Science, Western Sydney University, Locked Bag 1797, Penrith, NSW 2751, Australia}
\affiliation{CSIRO Astronomy and Space Science, PO Box 1130, Bentley, WA, 6102, Australia}

\author[0000-0003-4496-9553]{Amir Kazemi-Moridani}
\affiliation{Department of Physics and Astronomy, Rutgers, the State University of New Jersey, 136 Frelinghuysen Road, Piscataway, NJ 08854-8019, USA}

\author[0000-0003-3599-1521]{Bradley Frank}
\affiliation{South African Radio Astronomy Observatory, 2 Fir Street, Black River Park, Observatory, 7925, South Africa}
\affiliation{Inter-University Institute for Data Intensive Astronomy (IDIA)}
\affiliation{Department of Astronomy, University of Cape Town, Private Bag X3, Rondebosch 7700, South Africa}

\author[0000-0003-0046-9848]{Hayley Roberts}
\affiliation{Center for Astrophysics and Space Astronomy, Department of Astrophysical and Planetary Sciences, University of Colorado, 389 UCB, Boulder, CO 80309-0389, USA}

\author[0000-0003-2511-2060]{Jeremy Darling}
\affiliation{Center for Astrophysics and Space Astronomy, Department of Astrophysical and Planetary Sciences, University of Colorado, 389 UCB, Boulder, CO 80309-0389, USA}

\author[0000-0002-0648-2704]{Hans-Rainer Kl\"ockner}
\affiliation{Max-Planck-Institut f\"ur Radioastronomie, Auf dem H\"ugel 69, 53121, Bonn, Germany}

\author[0000-0003-4875-6272]{Nathan Adams}
\affiliation{Astrophysics, University of Oxford, Denys Wilkinson Building, Keble Road, Oxford OX1 3RH, UK}
\affiliation{Jodrell Bank Centre for Astrophysics, School of Physics and Astronomy, University of Manchester, Manchester M13 9PL, UK}

\author[0000-0002-7892-396X]{Andrew J. Baker}
\affiliation{Department of Physics and Astronomy, Rutgers, the State University of New Jersey, 136 Frelinghuysen Road, Piscataway, NJ 08854-8019, USA}
\affiliation{Department of Physics and Astronomy, University of the Western Cape, Robert Sobukwe Road, Bellville 7535, South Africa}

\author[0000-0002-3131-4374]{Matthew Bershady}
\affiliation{Department of Astronomy, University of Wisconsin-Madison, 475 N. Charter Street, Madison, WI 53706, USA}
\affiliation{South African Astronomical Observatory, P.O. Box 9, Observatory 7935, South Africa}
\affiliation{Department of Astronomy, University of Cape Town, Private Bag X3, Rondebosch 7700, South Africa}

\author[0000-0001-8404-848X]{Tariq Blecher}
\affiliation{Department of Physics and Electronics, Rhodes University, PO Box 94, Makhanda (Grahamstown), 6140, South Africa}

\author[0000-0002-5777-0036]{Sarah-Louise Blyth}
\affiliation{Department of Astronomy, University of Cape Town, Private Bag X3, Rondebosch 7700, South Africa}

\author[0000-0003-3917-1678]{Rebecca Bowler}
\affiliation{Astrophysics, University of Oxford, Denys Wilkinson Building, Keble Road, Oxford OX1 3RH, UK}
\affiliation{Jodrell Bank Centre for Astrophysics, School of Physics and Astronomy, University of Manchester, Manchester M13 9PL, UK}

\author[0000-0002-7625-562X]{Barbara Catinella}
\affiliation{International Centre for Radio Astronomy Research (ICRAR), The University of Western Australia, 35 Stirling Highway, Perth, WA 6009, Australia}
\affiliation{ARC Centre of Excellence for All Sky Astrophysics in 3 Dimensions (ASTRO 3D)}

\author[0000-0002-3834-7937]{Laurent Chemin}
\affiliation{Centro de Astronomia -- CITEVA, Universidad de Antofagasta, Avenida Angamos 601, Antofagasta 1270300, Chile}

\author[0000-0002-8969-5229]{Steven M. Crawford}
\affiliation{unaffiliated}

\author{Catherine Cress}
\affiliation{Department of Physics and Astronomy, University of the Western Cape, Robert Sobukwe Road, Bellville 7535, South Africa}

\author[0000-0003-2842-9434]{Romeel Dav\'e}
\affiliation{Institute for Astronomy, University of Edinburgh, Royal Observatory, Blackford Hill, Edinburgh, EH9 3HJ, UK}
\affiliation{Department of Physics and Astronomy, University of the Western Cape, Robert Sobukwe Road, Bellville 7535, South Africa}

\author[0000-0003-1027-5043]{Roger Deane}
\affiliation{Wits Centre for Astrophysics, School of Physics, University of the Witwatersrand, 1 Jan Smuts Avenue 2000, South Africa}
\affiliation{Department of Physics, University of Pretoria, Private Bag X20, Pretoria 0028, South Africa}

\author[0000-0001-8957-4518]{Erwin de Blok}
\affiliation{ASTRON, the Netherlands Institute for Radio Astronomy, Oude Hoogeveensedijk 4, 7991 PD, Dwingeloo, The Netherlands}
\affiliation{Department of Astronomy, University of Cape Town, Private Bag X3, Rondebosch 7700, South Africa}
\affiliation{Kapteyn Astronomical Institute, University of Groningen, P.O. Box 800, 9700 AV Groningen, The Netherlands}

\author[0000-0002-6149-0846]{Jacinta Delhaize}
\affiliation{Department of Astronomy, University of Cape Town, Private Bag X3, Rondebosch 7700, South Africa}

\author[0000-0001-6889-8388]{Kenneth Duncan}
\affiliation{Institute for Astronomy, University of Edinburgh, Royal Observatory, Blackford Hill, Edinburgh, EH9 3HJ, UK}

\author[0000-0001-9359-0713]{Ed Elson}
\affiliation{Department of Physics and Astronomy, University of the Western Cape, Robert Sobukwe Road, Bellville 7535, South Africa}

\author{Sean February}
\affiliation{South African Radio Astronomy Observatory, 2 Fir Street, Black River Park, Observatory, 7925, South Africa}

\author[0000-0003-1530-8713]{Eric Gawiser}
\affiliation{Department of Physics and Astronomy, Rutgers, the State University of New Jersey, 136 Frelinghuysen Road, Piscataway, NJ 08854-8019, USA}

\author[0000-0002-3065-457X]{Peter Hatfield}
\affiliation{Astrophysics, University of Oxford, Denys Wilkinson Building, Keble Road, Oxford OX1 3RH, UK}

\author[0000-0003-1020-8684]{Julia Healy}
\affiliation{ASTRON, the Netherlands Institute for Radio Astronomy, Oude Hoogeveensedijk 4, 7991 PD, Dwingeloo, The Netherlands}

\author[0000-0003-1530-8713]{Patricia Henning}
\affiliation{National Radio Astronomy Observatory, Pete V. Domenici Science Operations Center, P.O. Box O, 1003 Lopezville Road, Socorro, NM 87801-0387, USA}
\affiliation{Department of Physics and Astronomy, University of New Mexico, 210 Yale Boulevard NE, Albuquerque, NM 87106, USA}

\author[0000-0001-9662-9089]{Kelley M. Hess}
\affiliation{Instituto de Astrof\'{i}sica de Andaluc\'{i}a (CSIC), Glorieta de la Astronom\'{i}a s/n, 18008 Granada, Spain}
\affiliation{ASTRON, the Netherlands Institute for Radio Astronomy, Oude Hoogeveensedijk 4, 7991 PD, Dwingeloo, The Netherlands}
\affiliation{Kapteyn Astronomical Institute, University of Groningen, P.O. Box 800, 9700 AV Groningen, The Netherlands}

\author[0000-0001-6864-5057]{Ian Heywood}
\affiliation{Astrophysics, University of Oxford, Denys Wilkinson Building, Keble Road, Oxford OX1 3RH, UK}
\affiliation{Department of Physics and Electronics, Rhodes University, PO Box 94, Makhanda (Grahamstown), 6140, South Africa}
\affiliation{South African Radio Astronomy Observatory, 2 Fir Street, Black River Park, Observatory, 7925, South Africa}

\author[0000-0002-4884-6756]{Benne W. Holwerda}
\affiliation{Department of Physics and Astronomy, 102 Natural Science Building, University of Louisville, Louisville, KY 40292, USA}

\author[0000-0001-5449-143X]{Munira Hoosain}
\affiliation{Department of Astronomy, University of Cape Town, Private Bag X3, Rondebosch 7700, South Africa}

\author[0000-0002-8816-6800]{John P. Hughes}
\affiliation{Department of Physics and Astronomy, Rutgers, the State University of New Jersey, 136 Frelinghuysen Road, Piscataway, NJ 08854-8019, USA}

\author[0000-0002-8574-5495]{Zackary L. Hutchens}
\affiliation{Department of Physics \& Astronomy, CB3255, University of North Carolina, Chapel Hill, NC 27516, USA}

\author[0000-0001-7039-9078]{Matt Jarvis}
\affiliation{Astrophysics, University of Oxford, Denys Wilkinson Building, Keble Road, Oxford OX1 3RH, UK}
\affiliation{Department of Physics and Astronomy, University of the Western Cape, Robert Sobukwe Road, Bellville 7535, South Africa}

\author[0000-0002-3378-6551]{Sheila Kannappan}
\affiliation{Department of Physics \& Astronomy, CB3255, University of North Carolina, Chapel Hill, NC 27516, USA}

\author[0000-0002-3097-5381]{Neal Katz}
\affiliation{Department of Astronomy, University of Massachusetts, Amherst, MA 01003, USA}

\author{Du{\v s}an Kere{\v s}}
\affiliation{Center for Astrophysics and Space Sciences, University of California San Diego, 9500 Gilman Dr, La Jolla, CA 92093, USA}

\author[0000-0002-5882-610X]{Marie Korsaga}
\affiliation{Universit\'e de Strasbourg, CNRS, Observatoire astronomique de Strasbourg, UMR 7550, F-67000 Strasbourg, France}
\affiliation{Laboratoire de Physique et de Chimie de l’Environnement, Observatoire d’Astrophysique de l’Universit\'e Joseph Ki-Zerbo (ODAUO), 03 BP 7021,Ouaga 03, Burkina Faso}

\author[0000-0002-0202-6250]{Ren\'ee C. Kraan-Korteweg}
\affiliation{Department of Astronomy, University of Cape Town, Private Bag X3, Rondebosch 7700, South Africa}

\author[0000-0001-6841-6553]{Philip Lah}
\affiliation{Research School of Astronomy and Astrophysics, Australian National University, Canberra, ACT 2611, Australia}

\author[0000-0003-2221-8281]{Michelle Lochner}
\affiliation{Department of Physics and Astronomy, University of the Western Cape, Robert Sobukwe Road, Bellville 7535, South Africa}
\affiliation{South African Radio Astronomy Observatory, 2 Fir Street, Black River Park, Observatory, 7925, South Africa}

\author[0000-0001-8312-5260]{Natasha Maddox}
\affiliation{University Observatory, Faculty of Physics, Ludwig-Maximilians-Universit\"at, Scheinerstr. 1, 81679 Munich, Germany}

\author[0000-0001-9565-9622]{Sphesihle Makhathini}
\affiliation{Wits Centre for Astrophysics, School of Physics, University of the Witwatersrand, 1 Jan Smuts Avenue 2000, South Africa}

\author[0000-0002-0163-2507]{Gerhardt R. Meurer}
\affiliation{International Centre for Radio Astronomy Research (ICRAR), The University of Western Australia, 35 Stirling Highway, Perth, WA 6009, Australia}

\author[0000-0002-2838-3010]{Martin Meyer}
\affiliation{International Centre for Radio Astronomy Research (ICRAR), The University of Western Australia, 35 Stirling Highway, Perth, WA 6009, Australia}

\author[0000-0002-1527-0762]{Danail Obreschkow}
\affiliation{International Centre for Radio Astronomy Research (ICRAR), The University of Western Australia, 35 Stirling Highway, Perth, WA 6009, Australia}

\author[0000-0002-8379-0604]{Se-Heon Oh}
\affiliation{Department of Physics and Astronomy, Sejong University, 209 Neungdong-ro, Gwangjin-gu, Seoul, Republic of Korea}

\author[0000-0002-0616-6971]{Tom Oosterloo}
\affiliation{ASTRON, the Netherlands Institute for Radio Astronomy, Oude Hoogeveensedijk 4, 7991 PD, Dwingeloo, The Netherlands}
\affiliation{Kapteyn Astronomical Institute, University of Groningen, P.O. Box 800, 9700 AV Groningen, The Netherlands}

\author[0000-0003-2845-2714]{Joshua Oppor}
\affiliation{Department of Astronomy, University of Wisconsin-Madison, 475 N. Charter Street, Madison, WI 53706, USA}

\author[0000-0002-9160-391X]{Hengxing Pan}
\affiliation{Department of Physics and Astronomy, University of the Western Cape, Robert Sobukwe Road, Bellville 7535, South Africa}
\affiliation{Astrophysics, University of Oxford, Denys Wilkinson Building, Keble Road, Oxford OX1 3RH, UK}

\author[0000-0001-7996-7860]{D.~J. Pisano}
\affiliation{Department of Physics \& Astronomy, West Virginia University, Morgantown, WV 26506, USA}
\affiliation{Gravitational Wave and Cosmology Center, Chestnut Ridge Research Building, Morgantown, WV 26505, USA}

\author{Nandrianina Randriamiarinarivo}
\affiliation{Department of Physics and Astronomy, University of the Western Cape, Robert Sobukwe Road, Bellville 7535, South Africa}

\author[0000-0002-5269-6527]{Swara Ravindranath}
\affiliation{Space Telescope Science Institute, 3700 San Martin Drive, Baltimore, MD 21218-2410, USA}

\author[0000-0000-0000-0000]{Anja C. Schr\"oder}
\affiliation{Max-Planck-Institut f\"ur extraterrestrische Physik, Giessenbachstra{\ss}e 1, D-85748 Garching bei M\"unchen, Germany}

\author[0000-0001-7393-3336]{Rosalind Skelton}
\affiliation{South African Astronomical Observatory, P.O. Box 9, Observatory 7935, South Africa}

\author[0000-0003-1680-7936]{Oleg Smirnov}
\affiliation{Department of Physics and Electronics, Rhodes University, PO Box 94, Makhanda (Grahamstown), 6140, South Africa}
\affiliation{South African Radio Astronomy Observatory, 2 Fir Street, Black River Park, Observatory, 7925, South Africa}

\author[0000-0002-3321-1432]{Mathew Smith}
\affiliation{School of Physics and Astronomy, University of Southampton, Southampton, SO17 1BJ, UK}
\affiliation{Univ Lyon, Univ Claude Bernard Lyon 1, CNRS, IP2I Lyon / IN2P3, IMR 5822, F-69622, Villeurbanne, France}

\author{Rachel S. Somerville}
\affiliation{Center for Computational Astrophysics, Flatiron Institute, New York, NY 10010, USA}

\author[0000-0002-9062-1921]{Raghunathan Srianand}
\affiliation{IUCAA, Postbag 4, Ganeshkhind, Pune 411007, India}

\author[0000-0002-8057-0294]{Lister Staveley-Smith}
\affiliation{International Centre for Radio Astronomy Research (ICRAR), The University of Western Australia, 35 Stirling Highway, Perth, WA 6009, Australia}
\affiliation{ARC Centre of Excellence for All Sky Astrophysics in 3 Dimensions (ASTRO 3D)}

\author[0000-0002-5011-5178]{Masayuki Tanaka}
\affiliation{Department of Astronomical Science, The Graduate University for Advanced Studies, SOKENDAI, 2-21-1 Osawa, Mitaka, Tokyo, 181-8588, Japan}
\affiliation{National Astronomical Observatory of Japan, 2-21-1 Osawa, Mitaka, Tokyo, 181-8588, Japan}

\author[0000-0002-6748-0577]{Mattia Vaccari}
\affiliation{Inter-University Institute for Data Intensive Astronomy (IDIA)}
\affiliation{Department of Physics and Astronomy, University of the Western Cape, Robert Sobukwe Road, Bellville 7535, South Africa}
\affiliation{INAF - Istituto di Radioastronomia, via Gobetti 101, 40129 Bologna, Italy}

\author[0000-0003-4770-9829]{Wim van Driel}
\affiliation{GEPI, Observatoire de Paris, PSL Universit\'e, CNRS UMR 8111, 5 place Jules Janssen, 92190 Meudon, France}

\author[0000-0001-9022-8081]{Marc Verheijen}
\affiliation{Kapteyn Astronomical Institute, University of Groningen, Landleven 12, 9747 AD, Groningen, The Netherlands}

\author[0000-0003-4793-7880]{Fabian Walter}
\affiliation{Max-Planck-Institut f\"ur Astronomie (MPIA), K\"onigstuhl 16, 69117 Heidelberg, Germany}

\author[0000-0002-5077-881X]{John F. Wu}
\affiliation{Space Telescope Science Institute, 3700 San Martin Drive, Baltimore, MD 21218-2410, USA}

\author[0000-0003-0101-1804]{Martin A. Zwaan}
\affiliation{European Southern Observatory, Karl-Schwarzschild-Strasse 2, D-85748 Garching, Germany}



\begin{abstract}
In the local Universe, OH megamasers (OHMs) are detected almost exclusively in infrared-luminous galaxies, with a prevalence that increases with IR luminosity, suggesting that they trace gas-rich galaxy mergers. Given the proximity of the rest frequencies of OH and the hyperfine transition of neutral atomic hydrogen (H\,{\sc{i}}), radio surveys 
to probe the cosmic evolution of H\,{\sc{i}} in galaxies also offer exciting prospects for exploiting OHMs to probe the cosmic history of gas-rich mergers. Using observations for the Looking At the Distant Universe with the MeerKAT Array (LADUMA) deep H\,{\sc{i}} survey, we report the first untargeted detection of an OHM at $z > 0.5$, LADUMA\,J033046.20$-$275518.1 (nicknamed ``Nkalakatha''). The host system, WISEA\,J033046.26$-$275518.3, is an infrared-luminous radio galaxy whose optical redshift $z \approx 0.52$ confirms the MeerKAT emission line detection as OH at a redshift $z_{\rm OH} = 0.5225 \pm 0.0001$ rather than H\,{\sc{i}} at lower redshift. The detected spectral line has 18.4$\sigma$ peak significance, a width of $459 \pm 59\,{\rm km\,s^{-1}}$, and an integrated luminosity of $(6.31 \pm 0.18\,{\rm [statistical]}\,\pm 0.31\,{\rm [systematic]}) \times 10^3\,L_\odot$, placing it among the most luminous OHMs known. The galaxy's far-infrared luminosity $L_{\rm FIR} = (1.576 \pm 0.013) \times 10^{12}\,L_\odot$ marks it as an ultra-luminous infrared galaxy; its ratio of OH and infrared luminosities is similar to those for lower-redshift OHMs. A comparison between optical and OH redshifts offers a slight indication of an OH outflow. This detection represents the first step towards a systematic exploitation of OHMs as a tracer of galaxy growth at high redshifts.
 \end{abstract}
 


\section{Introduction} \label{sec:intro}

OH megamasers (OHMs) are luminous 18\,cm wavelength masers, produced in the centers of luminous and ultra-luminous infrared galaxies (LIRGs and ULIRGs) that have undergone merger-induced starburst activity \citep{2005ARA&A..43..625L}, are rich in dense gas \citep{2007ApJ...669L...9D}, and in some cases host luminous active galactic nuclei \citep[AGN;][]{2003Natur.421..821K}. Four 18\,cm lines connect the four hyperfine levels within the ${}^2\Pi_{3/2}(J = 3/2)$ ground state of the OH molecule: two main lines at 1665 and 1667\,MHz, and two satellite lines at 1612 and 1720\,MHz. In contrast to Galactic OH masers, extragalactic OHMs have large line widths and main line flux ratios $F_{1667}/F_{1665} > 1$. These attributes, along with the weakness of their satellite lines \citep{2013ApJ...774...35M}, can be naturally explained by a model in which OHMs are powered by radiative pumping through $53\,{\rm \mu m}$ lines that overlap in velocity, and in which different projected distributions of masing clumps can account for observations of both diffuse and compact emitting structures \citep{2008ApJ...677..985L}. 

The most extensive OHM survey in the local Universe has been conducted with Arecibo, with a focus on the detection of $z > 0.1$ systems \citep{2002AJ....124..100D}. In total, 53 OHMs were detected, spanning the redshift range $0.10 \leq z \leq 0.27$. Although \citet{2021A&A...648A.116C} report a tentative $3\sigma$ detection of the 1720\,MHz satellite line in emission in a targeted MeerKAT observation of the $z = 0.89$ quasar PKS\,1830$-$211, the {\it main} lines of OH have not been detected in emission at $z > 0.27$ up to now.

The local demographics of OHMs have been used to predict their occurrence at higher redshifts, where they are likely to represent a significant source of contamination for H\,{\sc{i}} surveys \citep{1998A&A...336..815B,2002ApJ...572..810D,2016MNRAS.459..220S,2021ApJ...911...38R}. Such surveys will be conducted by the interferometers that include the Square Kilometre Array (SKA) and its pathfinder facilities, such as MeerKAT \citep{2016mks..confE...1J} and the Australian SKA Pathfinder \citep[ASKAP;][]{2009IEEEP..97.1507D}. Although ULIRGs are more prevalent at higher redshifts \citep[][]{2005A&A...440L..17T} in part because normal star-forming galaxies at earlier epochs are likely to have $L_{\rm FIR} > 10^{12}\,L_\odot$ even in the absence of recent merging \citep[e.g.,][]{2008ApJS..175...48R}, OHMs have the potential to provide new constraints on the cosmic history of gas-rich mergers.

We report the first untargeted detection of an OHM at $z > 0.27$ from early observations for the Looking At the Distant Universe with the MeerKAT Array \citep[LADUMA;][]{2016mks..confE...4B} deep H\,{\sc{i}} survey. Section \ref{sec:obs} describes the acquisition and processing of the MeerKAT data in which the OH line was detected, and Section \ref{sec:res} presents our measurements and interpretation of the line parameters. In Section \ref{sec:disc}, we discuss the implications of this OHM in the context of previous knowledge of OHM hosts at lower redshifts and future observations probing to higher redshifts; \S \ref{sec:conc} summarizes our conclusions. The paper assumes a flat $\Lambda$CDM cosmology with $\Omega_m = 0.3$, $\Omega_\Lambda = 0.7$, and $H_0 = 73.3\,{\rm km\,s^{-1}\,Mpc^{-1}}$ \citep{2020MNRAS.498.1420W}.

\section{Observations} \label{sec:obs}

\subsection{MeerKAT} \label{subsec:mkat}

MeerKAT is a fixed-configuration array of 64 antennas equipped with receivers spanning the L (900--1670\,MHz) and UHF (580--1015\,MHz) bands. These overlapping frequency ranges access redshift ranges of $0 \leq z_{\rm H\,I} \leq 0.58$ and $0.40~\leq~z_{\rm H\,I}~\leq 1.45$ for the H\,{\sc{i}} line, and $0~\leq~z_{\rm OH}~\leq~0.85$ and $0.64~\leq~z_{\rm OH}~\leq~1.87$, adopting the stronger of the two main OH lines (at a rest frequency of 1667.359\,MHz) to define $z_{\rm OH}$.  The LADUMA deep H\,{\sc{i}} survey is using both L and UHF bands to probe the evolution of gas in galaxies over cosmic time. LADUMA is targeting a single pointing on the sky (03:32:30.4 $-$28:07:57 J2000) that encompasses the extended Chandra Deep Field South (ECDFS) and lies roughly at the  center of near-IR imaging coverage from the VISTA Deep Extragalactic Observations (VIDEO) survey \citep{2013MNRAS.428.1281J}.  Because the solid angle of MeerKAT's primary beam is inversely proportional to the square of the observing frequency, the cosmic volume LADUMA probes for any single spectral line expands with redshift 
like a trumpet (e.g., a South African vuvuzela).
Across the L band in particular, MeerKAT's circular field of view increases (at the half-power level) from $0.7\,{\rm deg}^2$ at 1667\,MHz to $0.9\,{\rm deg}^2$ at 1420.4\,MHz to $2.3\,{\rm deg}^2$ at 900\,MHz.

The first official L-band survey observation for LADUMA that used the 32k mode of the MeerKAT correlator (featuring 32,768 channels, each of width 26.1\,kHz) was taken on 2019 December 12 with 58 of the 64 antennas in operation. The bright radio galaxy PKS\,1934--63 was used as a flux and bandpass calibrator (observed for 10 minutes); the nearby quasar PKS\,0237--233 was 
observed as a gain calibrator for 3.5 minutes after each 20-minute observation of the LADUMA field. The total on-source integration time was 7.3 hours.

We reduced the data across the full L band, excluding regions affected by radio frequency interference (RFI), after splitting into 25\,MHz spectral windows (SPWs). For the source discussed in this paper, the relevant SPW spans 1086--1111\,MHz.  Bandpass, flux, and phase calibration were performed using the {\sc processMeerKAT} pipeline\footnote{\url{https://idia-pipelines.github.io/docs/processMeerKAT}}, which is written in Python, uses a purpose-built CASA \citep{2007ASPC..376..127M} Singularity container, and employs MPICASA (a parallelized form of CASA). Flux calibration used the \citet{reynolds1994} model for the spectrum of PKS\,1934$-$63, which is ultimately tied to northern hemisphere calibrators with flux uncertainties of $\sim 5\%$ at $\sim 1\,{\rm GHz}$ \citep[e.g.,][]{2017ApJS..230....7P}. The CASA task \textit{tclean} was used with robust~=~0 to create an initial continuum model as a basis for phase and amplitude self-calibration, after which model continuum visibility data were subtracted from the corrected visibility data using the CASA task  \textit{uvsub}. A third-order polynomial fit to the continuum was then calculated and subtracted using the CASA task \textit{uvcontsub} for all channels in each SPW to remove residual continuum emission from the spectral line data. Finally, spectral line cubes were created using \textit{tclean} with robust~=~0.5 and no cleaning; all channels in the resulting (dirty) cubes were convolved to a common synthesized beam of $14.2^{\prime\prime} \times 12.2^{\prime\prime}$ at a position angle of $-$15.8$^{\circ}$. The RMS per 26.1\,kHz channel was found to increase with frequency across the 1086--1111\,MHz SPW, with a value of 0.40~mJy\,beam$^{-1}$ near 1095\,MHz. All data were reduced on the ilifu cloud computing facility\footnote{\url{https://docs.ilifu.ac.za/\#/about/what_is}}. 

Visual inspection of the continuum-subtracted 26.1~kHz channel data cube with the Cube Analysis and Rendering Tool for Astronomy \citep[CARTA;][]{CARTA} revealed a bright spectral line at an observed frequency of 1095\,MHz, which we designate as LADUMA\,J033046.20$-$275518.1 and describe in detail in \S 3.1.

\begin{deluxetable*}{cccc}
\tablenum{1}
\tablecaption{Selected photometry for LADUMA\,J033046.20$-$275518.1. 
\label{tab:phot}}
\tablewidth{0.5pt}
\tablehead{
\colhead{ } \vspace{-0.05cm} & \colhead{Observed} & \colhead{Flux} & \colhead{ } \\
\colhead{Source} \vspace{-0.05cm} & \colhead{wavelength} & \colhead{density} & \colhead{Reference}
}
\startdata
WISE & $3.4\,{\rm \mu m}$ & $0.122 \pm 0.005\,{\rm mJy}$ & \citet{2010AJ....140.1868W} \\
WISE & $4.6\,{\rm \mu m}$ & $0.084 \pm 0.007\,{\rm mJy}$ & \citet{2010AJ....140.1868W} \\
WISE & $12\,{\rm \mu m}$ & $0.928 \pm 0.085\,{\rm mJy}$ & \citet{2010AJ....140.1868W} \\
{\it Spitzer}/MIPS & $70\,{\rm \mu m}$ & $141.11 \pm 0.66\,{\rm mJy}^a$ & \citet{2015ApJS..217...17H} \\
{\it Spitzer}/MIPS & $160\,{\rm \mu m}$ & $208.34 \pm 2.13\,{\rm mJy}^a$ & \citet{2015ApJS..217...17H} \\
{\it Herschel}/SPIRE & $250\,{\rm \mu m}$ & $90.93 \pm 0.65\,{\rm mJy}$ & \citet{2021MNRAS.507..129S} \\ 
{\it Herschel}/SPIRE & $350\,{\rm \mu m}$ & $42.36 \pm 1.53\,{\rm mJy}$ & \citet{2021MNRAS.507..129S} \\
{\it Herschel}/SPIRE & $500\,{\rm \mu m}$ & $17.6 \pm 3.7\,{\rm mJy}$ & \citet{2021MNRAS.507..129S} \\
ATCA  & 20\,cm & $0.42 \pm 0.029\,{\rm mJy}$ & \citet{2015MNRAS.453.4020F} \\
\enddata
\tablecomments{ ${}^{a}$ flux density extracted from a point spread function (PSF) fit}
\end{deluxetable*}

\subsection{Archival data} \label{subsec:arch}

LADUMA\,J033046.20$-$275518.1 has 
a clear galaxy counterpart in WISE imaging (WISEA\,J033046.26--275518.3) and in previous radio continuum mapping by the Australia Telescope Large Area Survey \citep{2012MNRAS.426.3334M,2015MNRAS.453.4020F}. The galaxy has a bent-tail radio morphology \citep{2014AJ....148...75D}, and has been classified as a narrow-angle (or head-tail) radio galaxy (that is, the bent radio structure lies on one side of the optical host galaxy).  The source has also been detected in dust emission by the Multiband Imaging Photometer for {\it Spitzer} \citep[MIPS;][]{2004ApJS..154...25R} on the {\it Spitzer Space Telescope} \citep{2015ApJS..217...17H} and by the Spectral and Photometric Imaging Receiver \citep[SPIRE;][]{2010A&A...518L...3G} on the {\it Herschel Space Observatory} \citep{2012MNRAS.424.1614O,2021MNRAS.507..129S}. Table~\ref{tab:phot} lists selected photometry for the source in the infrared and radio. Optical spectroscopy has yielded two independent redshift measurements of $z = 0.5245$ \citep{2009ApJ...707.1779E} and $z = 0.5247$ \citep{2012MNRAS.426.3334M}, neither with a quoted uncertainty; in this paper, we average these two measurements and adopt the $\sim 100\,{\rm km\,s^{-1}}$ uncertainty that is typical for AAT redshifts of similar vintage \citep[see, e.g., \S 1 of][]{2014MNRAS.441.2440B}, yielding $z_{\rm opt} = 0.5246 \pm 0.0005$.

We can determine the host galaxy's infrared luminosity using the source redshift and far-infrared photometry. By fitting emissivity-weighted blackbody functions to 1000 realizations of the {\it Spitzer}/MIPS and {\it Herschel}/SPIRE flux densities between 70--350\,$\mu$m (sampling their respective uncertainties), we derive a rest-frame dust temperature  
$T_d = 41.9 \pm 0.6\,{\rm K}$, a dust emissivity index $\beta = 1.67 \pm 0.06$ (for $\tau_\nu \propto \nu^\beta$), and a far-IR (rest-frame 42.5--122.5\,${\rm \mu m}$) flux $(6.30 \pm 0.05) \times 10^{-12}$\,erg\,s$^{-1}$\,cm$^{-2}$. Using the cosmology adopted for this paper and the redshift measured in \S 3.1 below, we then arrive at $L_{\rm FIR} = (1.576 \pm 0.013) \times 10^{12}\,L_{\odot}$, confirming the source as a ULIRG as previously noted by \citet{2011ApJ...727...83M}.

\begin{figure*}
\centering
\includegraphics[width=0.99\linewidth]{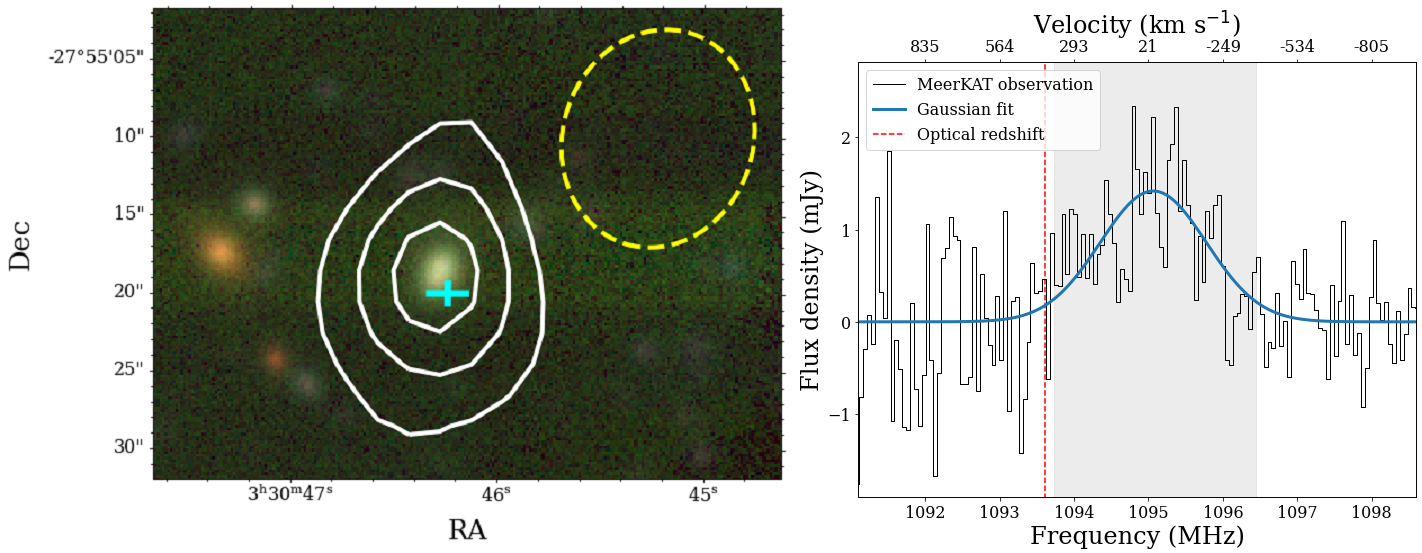}
\caption{\emph{Left:} RGB composite image made from the $grz$ filters of archival Hyper Suprime-Cam (HSC) data. White contours are from the zeroth moment map of our MeerKAT observation with no masking applied, integrated across $\sim 600\,{\rm km\,s^{-1}}$, at multiples of 5$\sigma$ (1$\sigma$~=~$0.0295\,{\rm Jy\,beam^{-1}\,km\,s^{-1}}$). The cyan cross indicates the centroid position and uncertainties from the \citet{2015MNRAS.453.4020F} observation with ATCA. The $14.2^{\prime\prime} \times 12.2^{\prime\prime}$ MeerKAT synthesized beam is plotted as a dashed yellow ellipse. \emph{Right:} Spectrum of the detected OHM, with frequency Doppler-corrected to a heliocentric reference frame and rebinned from its native 26.1~kHz resolution by a factor of 2. The rest-frame velocity range included in the moment map is shaded grey; the best Gaussian fit to the OH emission profile is overlaid in blue. The red dashed line indicates the frequency (1093.7\,MHz)} where the 1667\,MHz line would have appeared for $z_{\rm OH}$~=~$z_{\rm opt}$; that $z_{\rm OH} < z_{\rm opt}$ suggests a possible OH outflow.
\label{fig:overlayspect}
\end{figure*}

\begin{figure}
\centering
\includegraphics[width=0.99\linewidth]{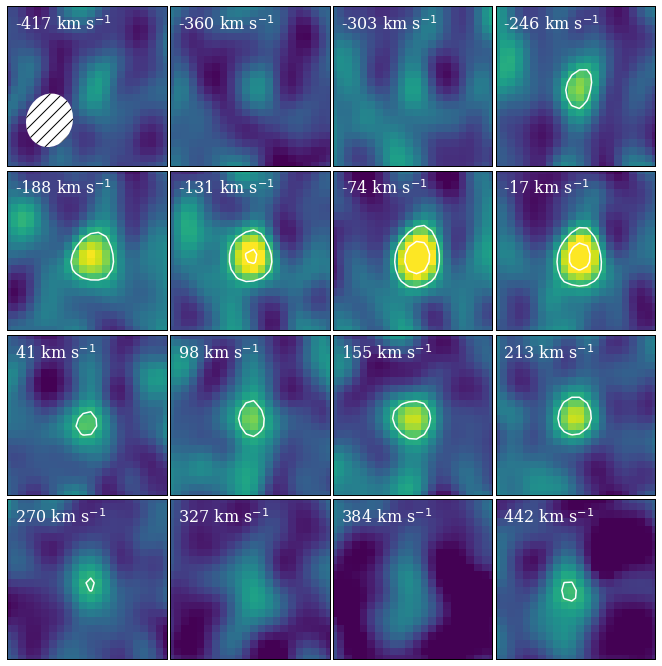}
\caption{Independent velocity channel maps of a 8$\times$ rebinned 26.1~kHz data cube, with velocities indicated relative to our measured systemic redshift of $z_{\rm OH} = 0.5225$. Contours are multiples of 2$\sigma$ (1$\sigma$~=~$0.283\,{\rm mJy\,beam^{-1}}$). The $14.2^{\prime\prime} \times 12.2^{\prime\prime}$ synthesized beam is plotted in the top left panel.}
\label{fig:channelslices}
\end{figure}

\section{Results} \label{sec:res}

\subsection{OH emission properties}

The spatial centroid of LADUMA\,J033046.20--275518.1 lies at right ascension 03:30:46.20 and declination $-$27:55:18.17 (J2000) with uncertainties of $\pm 0.7^{\prime\prime}$ (calculated as beam size divided by S/N) --- consistent with the centroids determined for its counterparts at other wavelengths, within their own uncertainties of $\sim 0.5^{\prime\prime}$ or greater. At this position and the observed line frequency, the primary beam correction was a factor $\sim (0.72)^{-1}$, which we applied using CASA before measuring line parameters.
  
In the right panel of Figure~\ref{fig:overlayspect}, we show the spectrum integrated over a $15^{\prime\prime}$ aperture, with frequency Doppler-corrected to the heliocentric reference frame and rebinned by a factor of 2. At the observed frequency, each rebinned (52.2\,kHz) channel corresponds to $14.3\,{\rm km\,s^{-1}}$ in rest-frame velocity. A single-Gaussian fit to the line spectrum yields (with fit uncertainties noted as errors) a central frequency of $\nu = 1095.07 \pm 0.09\,{\rm MHz}$, a rest-frame line width $\Delta v = 459 \pm 59\,{\rm km\,s^{-1}}$ (full-width at half-maximum, which corresponds to an observed frequency width of 1.68~MHz), a peak flux density of $1.42 \pm 0.16\,[\pm 0.07]$~mJy,\footnote{Errors in brackets represent 5\% flux scale uncertainties.} and a total spectral line flux $F_{\rm line} = 0.69 \pm 0.02\,[\pm 0.03]\,{\rm Jy\,km\,s^{-1}}$ that agrees well with a direct integral of the spectrum ($0.69 \pm 0.02\,[\pm 0.03]\,{\rm Jy\,km\,s^{-1}}$).

We note that the 1665\,MHz OH line may contribute to the measured line flux and width, although this fainter feature is not yet detected in this single-track observation, nor currently preferred over a single Gaussian fit. Given the presence of a multiwavelength counterpart with a previously measured optical redshift $z = 0.5246$, it is clear that our detection is hydroxyl emission at $z_{\rm OH} = 0.5225 \pm 0.0001$ (identified as the 1667.359~MHz transition) rather than  H\,{\sc{i}} emission at $z_{\rm H\,I} = 0.2970$. From the integrated line flux, we use the general relation for spectral line luminosity
\begin{equation}
{\frac {L_{\rm line}}{L_\odot}} = 1.0234\,\Big({\frac {\nu} {\rm MHz}}\Big)\,\Big({\frac {F_{\rm line}}{\rm Jy\,km\,s^{-1}}}\Big)\,\Big({\frac {D_L}{\rm Gpc}}\Big)^2
\end{equation}
in terms of observed frequency $\nu = 1095.1 \pm 0.1 \,{\rm MHz}$ and luminosity distance $D_L = 2.85 \pm 0.02 \,{\rm Gpc}$ for our adopted cosmology, obtaining an equivalent isotropic luminosity of 
$L_{\rm OH} = (6.31 \pm 0.18\,[\pm 0.31]) \times 10^3\,L_\odot$. 

The left panel of Figure~\ref{fig:overlayspect} shows contours from the zeroth moment map for the source, integrated over $600\,{\rm km\,s^{-1}}$ and overlaid on a composite image made from the $grz$ filters of archival Hyper Suprime-Cam \cite[HSC;][]{2018PASJ...70S...1M} data. The cyan cross indicates the ATCA position and its uncertainties from \citet{2015MNRAS.453.4020F}, and the yellow ellipse shows the MeerKAT synthesized beam. The OH emission is spatially unresolved, with S/N $\sim 18.4$ relative to the RMS in the zeroth moment map away from the source. Figure~\ref{fig:channelslices} shows the velocity channel maps of the source after rebinning by an additional factor of 4 to a rest-frame velocity resolution of $57.2\,{\rm km\,s^{-1}}$.

\subsection{Confirmation of the 1667~MHz identification}

\begin{figure*}
\centering
\includegraphics[width=0.99\linewidth]{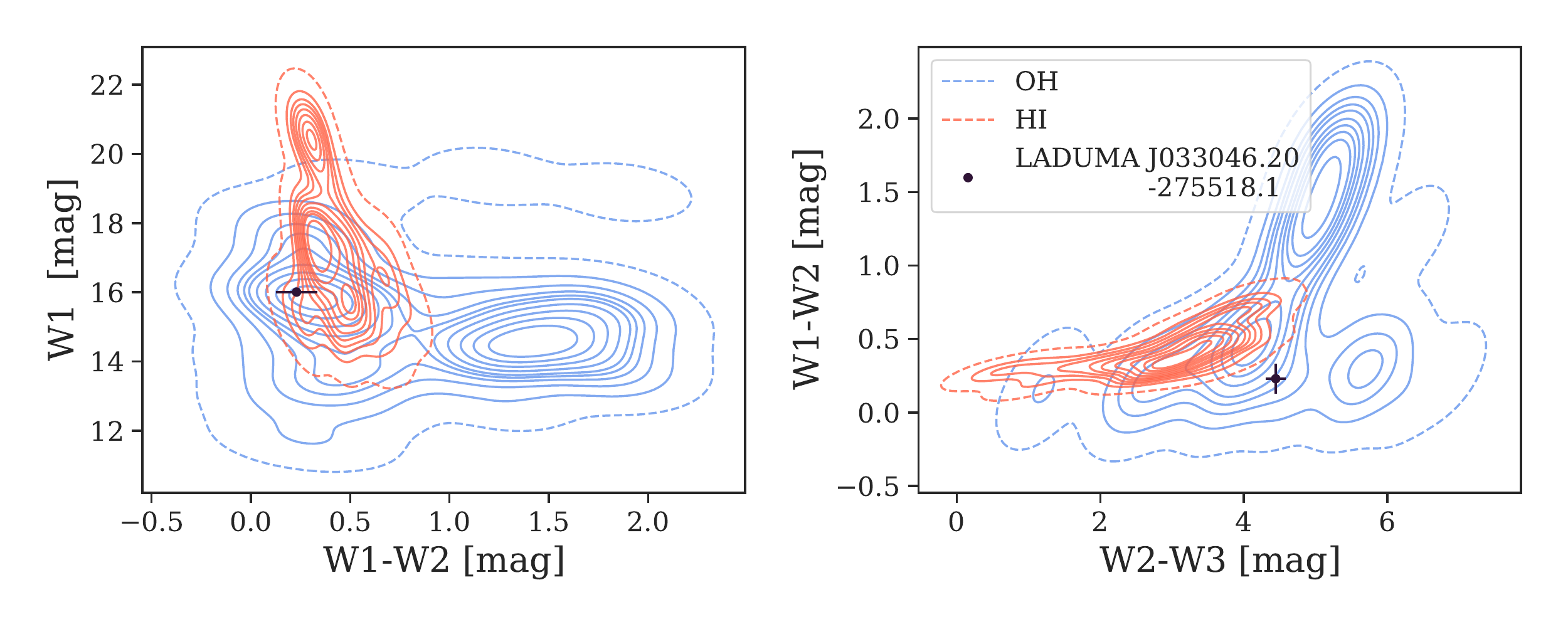}
\caption{Predicted distributions of gas-rich mergers traced by OH at $z_{\rm OH} \sim 0.52$ (blue), and H\,{\sc i}-rich disk galaxies at $z_{\rm H\,I} \sim 0.30$ (red), in WISE color-magnitude and color-color space, generated using algorithms presented in \citet{2021ApJ...911...38R}.  
Contours enclose the upper 99\% (dashed), 90\%, 80\%,... 20\%, and 10\% of the Gaussian kernel density estimates for the respective galaxy distributions.
In the left panel, both OH and H\,{\sc i} identifications are supported for LADUMA\,J033046.20$-$275518.1, while in the right panel, OH is clearly preferred.}
\label{fig:WISEcomp}
\end{figure*}

Notwithstanding the good agreement with a previously measured optical redshift, we have considered whether the detected emission could correspond to a weaker OH line \citep[e.g., the 1612 or 1720\,MHz satellite line, which can exhibit conjugate behavior:][]{2004ApJ...612...58D} rather than the main 1667\,MHz transition. In this scenario, we would expect an additional detection of the (brighter) 1667\,MHz main line at a different frequency (1132.5 or 1061.2~MHz) in our spectrum. No such emission feature is seen at the radio position. We also looked for emission in the satellite lines assuming our detection is indeed the 1667\,MHz main line at 1058.9 or 1130.0~MHz, with no such features seen. This result is not surprising given the weakness of satellite lines in OHMs \citep{2013ApJ...774...35M}, although as the LADUMA survey proceeds and reaches greater depths, we may be able to detect them in some systems. We also found no evidence of H{\sc i} emission or absorption at this redshift, although such features may become evident from the upcoming deeper LADUMA observations. 

In Figure~\ref{fig:WISEcomp}, we consider the implications of WISE magnitudes and colors for our identification of the detected emission line. We employ the algorithms presented in \citet{2021ApJ...911...38R}, who use machine learning to determine the redshift evolution in WISE magnitude and color space for an OHM host and a typical H{\sc i} source. The predicted distributions in WISE properties expected for gas-rich disk galaxies emitting H\,{\sc i} at redshift $z \sim 0.30$ (red) and a galaxy merger traced through OH at redshift $z \sim 0.52$ (blue) show that an OH identification for our source is plausible 
for all diagnostics, while an H\,{\sc i} identification is not always supported (see right panel). Therefore, we are confident that the emission detected corresponds to the main OH transition at 1667\,MHz.

\section{Discussion} \label{sec:disc}

\subsection{OH and FIR luminosities}

\begin{figure}
\centering
\includegraphics[width=0.99\linewidth]{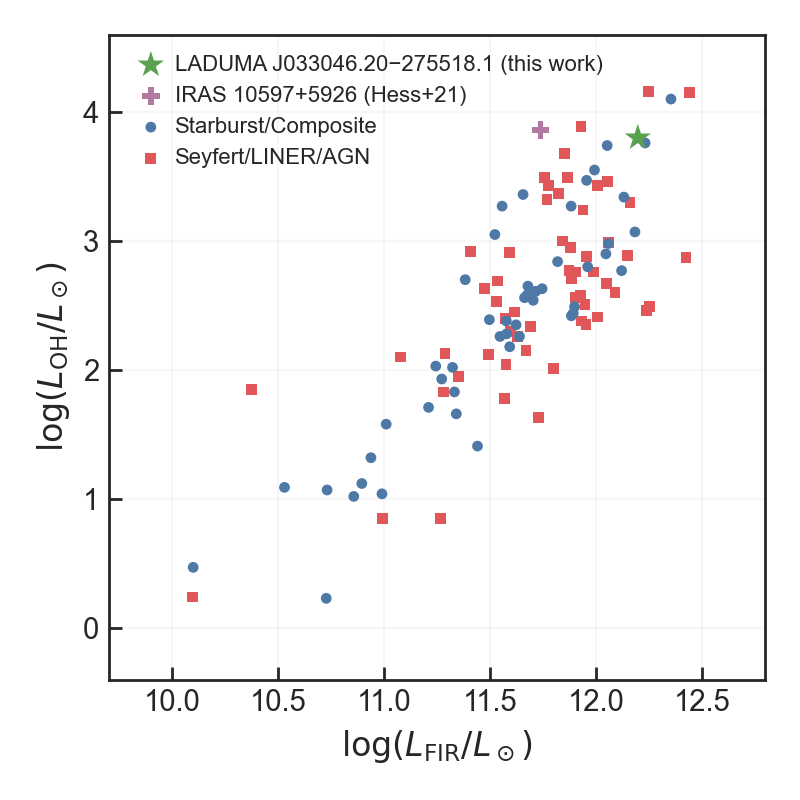}
\caption{$L_{\rm OH}$ vs. $L_{\rm FIR}$ for known OHMs (Roberts et al. 
2022, in preparation). LADUMA\,J033046.20$-$275518.1 (green star) is one of the most luminous OHMs on both axes.}
\label{fig:lumincomp}
\end{figure}

LADUMA\,J033046.20$-$275518.1 is one of the most luminous OHMs known at any redshift; only three sources within the sample of 53 OHMs presented in \cite{2002AJ....124..100D} and \cite{2002ApJ...572..810D} have higher $L_{\rm OH}$ values, and it is only marginally less luminous than the $z = 0.1961$ OHM recently detected by Apertif \citep{2021A&A...647A.193H}, which would have $L_{\rm OH} = 7.18 \times 10^{3}\,L_{\odot}$ for our cosmology. In recognition of this power and its unprecedentedly high redshift, we have given it the nickname ``Nkalakatha,'' an isiZulu word that means ``big boss.'' It is not overly surprising that a detection reached in a single LADUMA track will be among the most luminous OHMs when compared to sources from the local Universe at $z < 0.27$. Using the relationship between OH and far-IR luminosities derived by \cite{2002AJ....124..100D} for an Arecibo survey + literature sample 
of OHMs, i.e., 
${\rm log}\,(L_{\rm OH}/L_\odot) = (1.57 \pm 0.11)\,{\rm log}\,(L_{\rm FIR}/L_\odot) - (15.76 \pm 1.22)$,
we find that the predicted OH luminosity for LADUMA\,J033046.20$-$275518.1 is $L_{\rm OH pred} \approx 2.4 \times 10^{3}\,L_{\odot}$. This is roughly a factor of 2.6 smaller than what we observe, but well within the large scatter observed for the local relation (Fig.~\ref{fig:lumincomp}). The dust temperature recovered from the fit to the system's FIR photometry ($T_d \approx 42\,{\rm K}$) 
is roughly consistent with theoretical expectations and observational results that $T_d \geq 45\,{\rm K}$ is required for OH masing to occur \citep{2008ApJ...677..985L,2011ApJ...730...56W}, particularly if cooler dust outside the masing region contributes to $L_{\rm FIR}$, and is at the lower end of the distribution of global dust temperatures for OHM hosts.

We have also considered the FIR–radio flux ratio parameter $q$, defined by \cite{1985ApJ...298L...7H} as
\begin{equation}
    q = {\rm log}\,\Big\{\Big({\frac{F_{\rm FIR}} {\rm W\,m^{-2}}}\Big)\,\Big({\frac {10^{29}}{3.75 \times 10^{12}\,{\rm Hz}}}\Big)\,\Big({\frac {S_{\rm 1.4\,GHz}}{\rm mJy}}\Big)^{-1}\,\Big\}
\end{equation}
For LADUMA\,J033046.20$-$275518.1, we correct the observed radio continuum flux density\footnote{This value is comparable to the value measured from MeerKAT continuum imaging of the LADUMA field, which will be published in a forthcoming paper analogous to \cite{2022MNRAS.509.2150H}.} (Table \ref{tab:phot}) to 1.4\,GHz in the rest frame assuming a spectrum $S_\nu \propto \nu^{-0.7}$, and arrive at an estimated $q \approx 2.7$. 
This value is consistent with the infrared-radio correlation for star-forming galaxies with no evidence of AGN \citep[see, e.g., Fig. 16 of][]{2017A&A...602A...4D}, higher than the median $q = 2.37$ recently measured for 89 star-forming galaxies in the COSMOS field \citep{2021MNRAS.507.2643A}, and much higher than the threshold $q \approx 1.8$ at which galaxies are three times more radio-loud than the mean for star-forming systems \citep{2002AJ....124..675C}. 
{\it Spitzer}/IRAC photometry for this source \citep{2003PASP..115..897L} also disfavor an AGN identification according to well-established criteria \citep{2005ApJ...631..163S,2012ApJ...748..142D}, as does a classification of its optical spectrum \citep{2012MNRAS.426.3334M}.
As a 
result, we view the far-IR emission from LADUMA\,J033046.20$-$275518.1 
as more likely due to starburst activity than to a bolometrically 
significant AGN. Such a conclusion is also not surprising given 
evidence from the local Universe that ULIRGs hosting OHMs are less 
likely to show evidence of AGN at infrared wavelengths than non-masing 
ULIRGs \citep{2011ApJ...730...56W}. 

\subsection{A possible molecular outflow}

Large-scale outflows emanating from the central regions of galaxies and powered by both starbursts and AGN have been known for decades to play an important role in the evolution of galaxies and the intergalactic medium \cite[e.g.,][]{1990ApJS...74..833H}. Such outflows have been detected through OH observations in a number of systems. \cite{1989ApJ...346..680B} detect three distinct outflows in emission in a sample of five OHMs, with one galaxy showing a maximum outflow velocity of 800\,km\,s$^{-1}$. \cite{2014A&A...561A..27G} meanwhile find far-IR OH features blueshifted by over 1000\,km\,s$^{-1}$ in Mrk~231, with the central AGN likely responsible for the high mass outflow rate and outflow velocities detected. 

The OH emission in LADUMA\,J033046.20$-$275518.1 is blueshifted by $407 \pm 118\,{\rm km\,s^{-1}}$ relative to the redshift measured from optical spectroscopy.
The simplest explanation for a one-sided velocity offset is a starburst-driven outflow, given the high value of $q$ noted above. However, the situation could be more complex, involving multiple nuclei or a disk close to an (obscured) AGN, possibly interacting with outflows.
As our MeerKAT observation of this galaxy and its OH emission is unresolved, higher spatial resolution observations would be required to further characterize the outflow's extent and energetics. 
The study of \cite{2018ApJ...859...35G}, investigating molecular gas outflows in two starburst ULIRGs in mid-$J$ CO and 18\,cm OH lines, is instructive in this regard. CO outflow velocities are seen to exceed 1600\,km\,s$^{-1}$, with corresponding mass outflow rates of 300--700\,$M_{\odot}$\,yr$^{-1}$; meanwhile, OH outflow velocities are seen to extend to 1000~km\,s$^{-1}$, with velocity wings for one source agreeing ``remarkably well'' with previous detections of outflowing molecular gas.
A higher-resolution study of other molecular species (e.g., CO) in LADUMA\,J033046.20$-$275518.1 may thus be a productive strategy for confirming the existence of an outflow in OH. 

\section{Conclusions} \label{sec:conc}

We present the first detection of an OHM in the LADUMA field, LADUMA\,J033046.20$-$275518.1 ``Nkalakatha,'' which is also the highest redshift detection of such a system to date in the main 1667\,MHz OH emission line. 
OH emission is found to be redshifted to $z_{\rm OH} = 0.5225$, which agrees well with an optical redshift of $z_{\rm opt} = 0.5246$ for a host galaxy already known to be a ULIRG. The system's total OH luminosity of $L_{\rm OH} = (6.31 \pm 0.18\,[\pm 0.31]) \times 10^3\,L_\odot$ makes it one of the most luminous OHMs known (all other 1667\,MHz detections have redshifts $z < 0.27$), and is consistent with its large far-IR luminosity. The $\sim$400~km\,s$^{-1}$ offset between the OH and optical redshifts is most simply explained by a starburst-driven outflow.

This detection highlights the potential of upcoming spectral line surveys, 
whose wide frequency coverage will enable further high-redshift measurements. \cite{2021ApJ...911...38R} predict that $83 \pm 20$ OHMs will be detected in LADUMA alone, which will nearly double the number of known OHMs.
Greater numbers are expected in lower redshift but wider area surveys, such as the Widefield ASKAP L-band Legacy All-sky Blind surveY \cite[WALLABY;][]{Koribalski2020}, the Apertif Wide-area Extragalactic Survey (AWES; Hess et al., in prep), and the H\,{\sc i} component \citep[MIGHTEE-H\,{\sc{i}};][]{Maddox2021} of the MeerKAT International GigaHertz Tiered Extragalactic Exploration (MIGHTEE) survey \citep{Jarvis2016}.
As LADUMA reaches greater depths, we expect to set further OHM redshift records, which will enable studies of the cosmic rate of gas-rich galaxy mergers and further constrain galaxy evolution models.


\acknowledgments

The authors sincerely thank Zolile Tibane for suggesting the nickname ``Nkalakatha'' for LADUMA\,J033046.20$-$275518.1. We also thank the anonymous referee for comments that have helped improve the paper.

The MeerKAT telescope is operated by the South African Radio Astronomy Observatory (SARAO; \url{www.sarao.ac.za}), which is a facility of the National Research Foundation (NRF), an agency of the Department of Science and Innovation. The authors thank the members of the SARAO engineering, commissioning, and science data processing teams for building and operating an absolutely superb facility. The MeerKAT data presented in this paper were processed using the ilifu cloud computing facility  (\url{www.ilifu.ac.za}), which is operated by a consortium that includes the University of Cape Town (UCT), the University of the Western Cape, the University of Stellenbosch, Sol Plaatje University, the Cape Peninsula University of Technology and the South African Radio Astronomy Observatory. The ilifu facility is supported by contributions from the Inter-University Institute for Data Intensive Astronomy (IDIA, which is a partnership between the UCT, the University of Pretoria and the University of the Western Cape), the Computational Biology division at UCT, and the Data Intensive Research Initiative of South Africa (DIRISA). Data processing used pipelines that were developed at IDIA and are available at \url{https://idia-pipelines.github.io}.

MG acknowledges support from IDIA and was partially supported by the Australian Government through the Australian Research Council's Discovery Projects funding scheme (DP210102103). AKM and AJB acknowledge support from NSF grant AST-1814421. AKM also thanks the LSSTC Data Science Fellowship Program, which is funded by the LSST Corporation, NSF grant OAC-1829740, the Brinson Foundation, and the Moore Foundation; his participation in the program has been helpful for this work. HR and JeD acknowledge support from NSF grant AST-1814648; MB and JO acknowledge support from NSF grant AST-1814682; RB acknowledges support from an STFC Ernest Rutherford Fellowship (grant number ST/T003596/1); JaD and HP acknowledge the financial assistance of SARAO; KMH acknowledges funding from the State Agency for Research of the Spanish Ministry of Science, Innovation and Universities through the ``Center of Excellence Severo Ochoa'' awarded to the Instituto de Astrof\'isica de Andaluc\'ia (SEV-2017-0709); from grant RTI2018-096228-B-C31 (Ministry of Science, Innovation and Universities / State Agency for Research / European Regional Development Funds, European Union); and from the coordination of the participation in SKA-SPAIN, funded by the Ministry of Science and innovation (MICIN); and ZLH and SK acknowledge support from NSF grant AST-1814486. Parts of this research were supported by the Australian Research Council Centre of Excellence for All Sky Astrophysics in 3 Dimensions (ASTRO 3D), through project number CE170100013, and by the South African Research Chairs Initiative of the Department of Science and Technology and the NRF.

%

\vspace{5mm}
\facilities{MeerKAT}






\bibliography{glowacki20}{}
\bibliographystyle{aasjournal}



\end{document}